\documentclass[aps,prx,twocolumn,preprintnumbers,amsmath,amssymb]{revtex4}

\usepackage{graphicx}
\usepackage{dcolumn}
\usepackage{bm}
\usepackage{mathrsfs}
\usepackage{subfigure}
\usepackage{natbib}
\usepackage{amsmath}
\usepackage{amssymb}
\usepackage{Jonasmacros}
\usepackage[usenames,dvipsnames,svgnames]{xcolor}
\usepackage{mathptmx}
\usepackage{txfonts}

\begin{document}
	\title{Transient spin dynamics in a single-molecule magnet}

	\author{H. Hammar}
	\affiliation{Department of Physics and Astronomy, Uppsala University, Box 530, SE-751 21 Uppsala}

	\author{J. Fransson}
	\affiliation{Department of Physics and Astronomy, Uppsala University, Box 530, SE-751 21 Uppsala}

	\date{\today}

\begin{abstract}
	We explore the limitations and validity of semi-classically formulated spin equations of motion.
	Using a single-molecule magnet as a test model, we employ three qualitatively different approximation schemes.
	From a microscopic model, we derive a generalized spin equation of motion in which the parameters have a non-local time-dependence.
	This dynamical equation is simplified to the Landau-Lifshitz-Gilbert equation with i) time-dependent, and ii) time-independent parameters.
	We show that transient dynamics is essentially non-existing in the latter approximation, while the former breaks down in the regime of strong coupling between the spin and the itinerant electrons.
\end{abstract}

\maketitle

\section{Introduction}
Single-molecule magnets have been of interest as the intrinsic spin moment of magnetic molecules makes them suitable for logical operations, and serve as good model systems to study fundamental physical phenomena \cite{Bogani:2008aa,Locatelli:2014aa,Chappert:2007aa}.
Experimentally it has been shown that one can control the magnetic moment and detect the spin excitations of molecules by electrical current \cite{Hauptmann:2008aa,Loth:2010aa,Wagner:2013aa, Vincent2012}.
Together with other new methods for probing single-molecule spin states \cite{Loth2010b,Ternes2015,Loth2010,Vincent2012,PhysRevLett.116.027201, PhysRevLett.115.206803}, control and read-out of single molecules and atoms is possible.
Experiments on single magnetic atoms and molecules show distance dependent effects in their exchange \cite{Hirjibehedin19052006,PhysRevLett.98.056601,Zhou:2010aa,Meier04042008},
large anisotropy of individual molecules  \cite{Rau2014,PhysRevLett.114.247203,PhysRevB.78.155403,PhysRevLett.102.257203}, as well as collective spin excitations and Kondo effect \cite{PhysRevLett.101.197208,PhysRevLett.103.107203,Pruser:2011aa,Khajetoorians04012013}. 
Recent experimental progress show long-time stability of the spin state of individual atoms on a surface \cite{Donati2016, Natterer2017, Paul2017}.
This, and other experiments
\cite{Mannini:2009aa,Mannini:2010aa,PhysRevLett.97.207201,Khajetoorians296,Khajetoorians27052011,Loth13012012}, open ways towards realization of single-atom memory devices.

Theoretically, one common approach for describing the dynamics of the magnetization in materials is to employ the phenomenological Landau-Lifshitz-Gilbert (LLG) equation of motion \cite{Ralph20081190}.
This has successfully been applied to describe the magnetization dynamics of different materials \cite{Ralph20081190}.
The LLG equation has been extended to take into account temperature, moment of inertia, and stochastic forces \cite{Evans2014, Ellis2015, PhysRevB.92.104403,PhysRevB.92.020410,Arrachea:2015aa,PhysRevE.92.012116,Ritzmann2014,Xiao2010}.
Due to the large interest in the field of ultra-fast spin dynamics \cite{Walowski2016}, further investigations has been done of the LLG equation in the ultra-fast regime \cite{Ellis2015,Evans2014} and on dynamic exchange interactions \cite{Secchi2013,Secchi2016,Mikhaylovskiy:2015aa,Mueller2013,Mentink2014}.

In this article we focus on the description of the spin dynamics of single-molecule magnets.
Methods using quantum master equations  \cite{PhysRevB.73.235304,PhysRevB.75.134425,PhysRevB.86.245317,PhysRevB.90.134305} and stochastic LLG equation \cite{PhysRevB.87.045426,PhysRevB.85.115440} have been thoroughly investigated.
Another technique, which will be used in this paper, is to derive a spin equation of motion (SEOM) from the spin action defined on the Keldysh contour, considering the nonequilibrium properties of the effective spin moment \cite{PhysRevB.77.205316,0953-8984-24-11-116001,PhysRevB.77.054401,PhysRevB.73.212501,0957-4484-19-28-285714}.
This provides a general description of the spin dynamics and exchange interactions in the nonequilibrium regime \cite{PhysRevLett.92.107001,PhysRevLett.113.257201,PhysRevLett.108.057204,PhysRevB.82.180411,1367-2630-10-1-013017}.
We remark, however, that while the employed approach holds well for localized spins in an electronic environment, it is not clear whether it is applicable to itinerant magnetism. Therefore, we restrict our discussions to localized spins only, e.g., M-phthalocyanines and M-porphyrins, where the transition metal $d$-levels, which are deeply localized, constitute the localized magnetic moment.
Similar approaches have previously been used in order to understand effects in the stationary limit, such as voltage dependence, geometric phases and chaotic behaviors \cite{PhysRevB.75.214420,PhysRevLett.114.176806,PhysRevLett.96.066603,PhysRevB.73.212501,Saygun2016a}.
Here, we study the transient regime, considered through a generalized SEOM where the parameters depend on both time and history.

\begin{figure}[b]
	\centering
	\includegraphics[width =0.7\columnwidth]{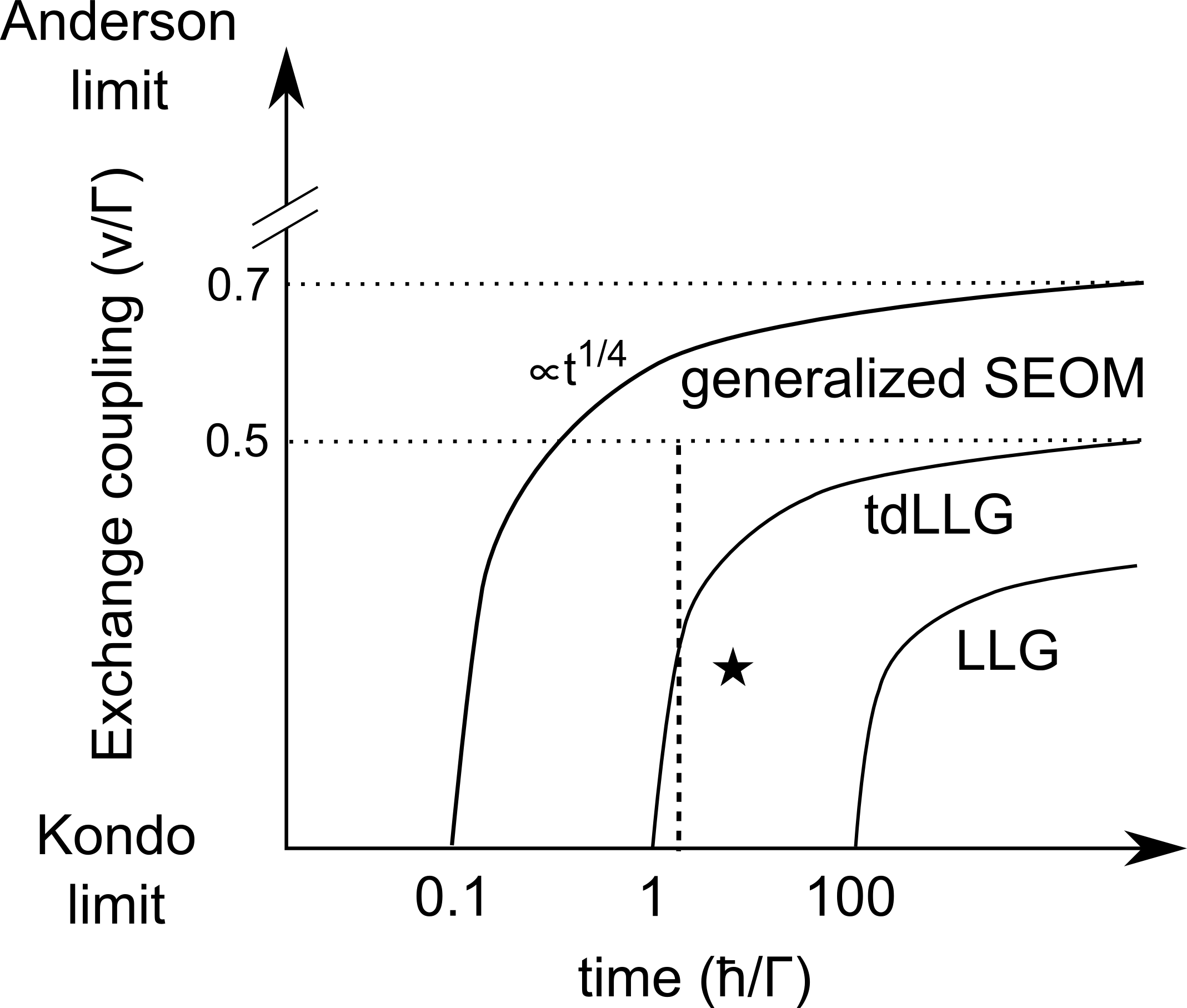}
	\caption{Diagram showing regimes of validity for the different approximation schemes used in this paper. The regimes scale with time t as $t^{1/4}$. Going from slower to faster time-scales, and from low to high exchange coupling, one needs to extend the LLG equation to incorporate quantum effects. Here, the generalized SEOM denotes the general approach used in this paper, and tdLLG denotes a LLG equation with time-dependent parameters. The horizontal dotted lines indicates the limits of the exchange coupling in terms of the model parameters. The star indicates the parameters of Fig. \ref{DiffSzsolution} while the vertical dotted line indicates the results in Fig. \ref{DiffExchange}.}
	\label{solutiondiagram}
\end{figure}

We examine the limitations of the LLG equation by comparing three different approximation schemes.
First, by making use of the Born-Oppenheimer approximation, one can derive a generalized SEOM where the parameters evolve with time and depends on the full memory of the system.
The second approximation scheme is to assume a slowly varying spin, such that we can disregard the spin history and retain a LLG equation with time-dependent parameters, henceforth referred to as tdLLG.
The third approximation scheme is obtained by considering the parameters of the generalized SEOM in the stationary limit, resulting into a LLG equation with constant parameters. This is the commonly used approach when performing LLG calculations.

Our main results are summarized in Fig. \ref{solutiondiagram} where the tunneling coupling, $\Gamma$, indicates the energy scales of the system.
The tunneling coupling also determines the time-scale of the memory of the system.
For short time-scales $t\ll0.1\hbar/\Gamma$, e.g., hundreds of femtoseconds for an exchange coupling of 1 meV, the electron dynamics become increasingly important and the Born-Oppenheimer approximation is no longer valid.
Therefore, the generalized SEOM is not sufficient and a full quantum mechanical treatment is necessary.
This is also true in the Anderson limit, where strong correlations have to be considered for the on-site terms.
The generalized SEOM is valid above 0.1 $\hbar/\Gamma$ and in systems which can be described by the Kondo model, e.g., a localized spin pertaining to magnetic molecules, with an exchange coupling smaller than 0.7 $\Gamma$.
The validity scales as $t^{1/4}$ as the dynamics scale with the exchange coupling, $v$, as $v^4$, see Fig. \ref{DiffExchange}.
Disregarding history of the spin, as in tdLLG, one can treat slower dynamics and more weakly coupled systems. 
This would be dynamics slower than 1 $\hbar/\Gamma$,  e.g., picoseconds for an exchange coupling of 1 meV, and an exchange coupling smaller than 0.5 $\Gamma$.
For slow dynamics ($\hbar\Gamma\gg1$) where the memory effects are negligible and with small or adiabatic changes, the Markovian approximation is justified and it suffices to use constant parameters in the LLG equation.
This approach fails though to account for rapid changes in the system, as shown in this paper.

Although our focus is on a single-molecule magnet, it is only an example of the general framework described in this article. The spin dynamics formalism introduced in the theory part of this article remains general for any system that has macroscopical spins. Similar treatments have been investigated in strongly correlated materials \cite{Secchi2013,Secchi2016}. We believe that our results have implications in larger nanostructures, as it shows the importance of a more inclusive description to incorporate rapid changes in the system. This is specially relevant with the current interest in ultra-fast spin dynamics.

The article is organized as follows.
In Sec. II, the theoretical background is introduced.
This includes the simple system studied, a discussion of the spin equation of motion, the exchange coupling and the stationary limit.
In Sec. III the results are presented and discussed, and the article is concluded in Sec. V.

\section{Theory}

\begin{figure}[t]
	\centering
	\includegraphics[width=\columnwidth]{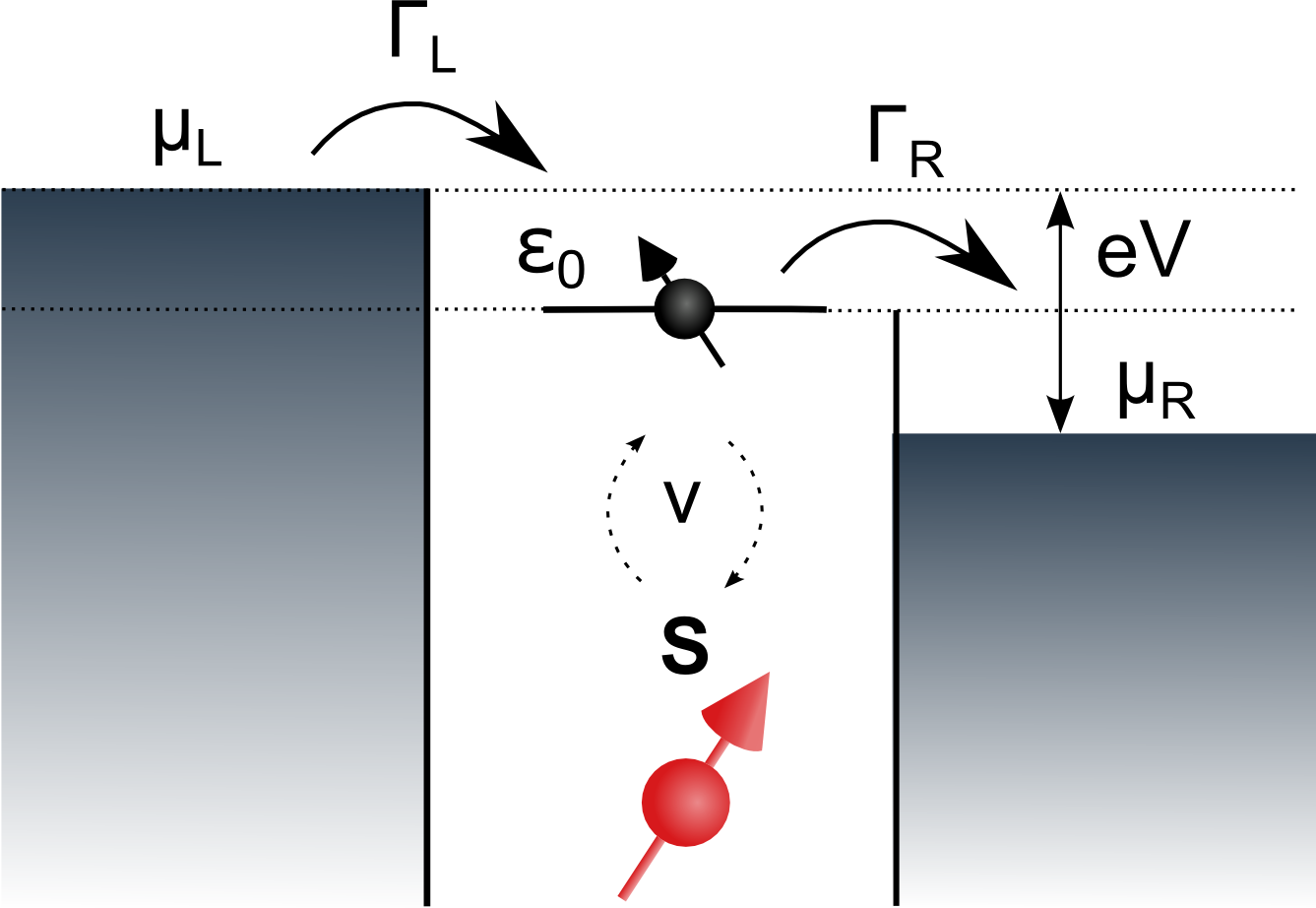}
	\caption{The system studied in this work consisting of a local magnetic moment coupled to a quantum dot in a tunnel junction between non-magnetic leads.}
	\label{system}
\end{figure}

\subsection{Model system}
We start by defining our model system. We consider a magnetic molecule embedded in a tunnel junction between metallic leads, see Ref. \cite{Hammar2016} and Fig. \ref{system} for reference.
The magnetic molecule comprises a localized magnetic moment ${\bf S}$ coupled via exchange to the highest occupied molecular orbital (HOMO) or lowest unoccupied molecular orbital (LUMO) level, henceforth referred to as the quantum dot level.

We define our system Hamiltonian as
\begin{equation}
{\cal H}
=
{\cal H_{{\rm \chi}}}+{\cal H_{{\rm T}}}+{\cal H_{{\rm QD}}}+{\cal H_{{\rm S}}}
\end{equation}
Here,
\begin{equation}
{\cal H}_{\chi}=\sum_{\bfk\sigma\in\chi} (\varepsilon_{\bfk\chi}-\mu_{\chi})c_{\bfk\chi\sigma}^{\text{\ensuremath{\dagger}}}c_{\bfk\chi\sigma}
\end{equation}
is the Hamiltonian for the lead $\chi = L/R$, where $c_{\bfk\chi\sigma}^{\dagger}$ ($c_{\bfk\chi\sigma}$) creates (annihilates) an electron in the lead with energy $\varepsilon_{\bfk\chi}$, momentum \textbf{k} and spin $\sigma=\up,\down$.
We have introduced the chemical potential $\mu_{\chi}$ for the leads and the voltage bias V across the junction defined as $eV = \mu_{L}-\mu_{R}$.
Each lead has the same temperature T.
Tunneling between the leads and the quantum dot level is described by ${\cal H}_{T}={\cal H}_{TL}+{\cal H}_{TR}$, where
\begin{equation}
{\cal H}_{T\chi}=T_{\chi}\sum_{\bfk\sigma\in\chi}c_{\mathbf{k\chi\sigma}}^{\dagger}d_{\sigma}+H.c.
\end{equation}
Using the wide-band limit we can define the tunneling coupling $\Gamma^\chi=2|T_\chi|^2\sum_{\bfk\in\chi}\delta(\omega-\varepsilon_{\bfk})$ between the lead and the quantum dot.
The single-level quantum dot is represented by ${\cal H}_{QD}=\sum_{\sigma}\varepsilon_{\sigma}d_{\sigma}^{\dagger}d_{\sigma}$, where $d_{\sigma}^{\dagger}$ ($d_{\sigma}$) creates (annihilates) an electron in the quantum dot with energy $\varepsilon_{\sigma} = \varepsilon_{0} + g\mu_{B} B \sigma^{z}_{\sigma\sigma}/2$ and spin $\sigma$.
We include the Zeeman split due to the external magnetic field $\bfB=B\hat{\bf z}$, where $g = 2$ is the gyromagnetic ratio and $\mu_{B}$ the Bohr magneton.
The local spin is described by
\begin{equation}
{\cal H}_{\rm S}=-g\mu_B\bfS\cdot\bfB-v\mathbf{s}\cdot\mathbf{S}.\label{spinHamiltonian}
\end{equation}
Here,$v$ is the exchange integral between the localized and delocalized electrons, the electron spin is denoted \textbf{$\mathbf{s}=\psi^{\dagger}\boldsymbol{\sigma}\psi/2$}, defined in terms of the spinor $\psi=(d_\up\ d_\down)^t$, and where $\boldsymbol{\sigma}$ is the vector of Pauli matrices.

We introduce a contour ordered Green's function ${\bf G} (t,t')$ for the electrons in the quantum dot. The lesser and greater matrix Green's function is defined as ${\bf G}^<(t,t')=\{i\langle d^\dagger_{\sigma'} (t')d_\sigma(t)\rangle\}_{\sigma\sigma'}$ and ${\bf G}^>(t,t')=\{(-i)\langle d_\sigma(t)d^\dagger_{\sigma'} (t')\rangle\}_{\sigma\sigma'}$ respectively. As we are interested in the response of the spin dynamics in the current flowing through the molecule we make a first order expansion of the Green's function with respect to the spin. In the case of non-magnetic leads and vanishing external magnetic field the contour ordered Green's function takes the form
\begin{align}
\mathbf{G} (t,t')=&
\textbf{g} (t,t')-v\oint_C \textbf{g} (t,\tau)\left\langle \mathbf{S} (\tau)\right\rangle \mathbf{\cdot\boldsymbol{\sigma}}\textbf{g} (\tau,t')d\tau.
\end{align}
Here, $\textbf{g} (t,t')$ is the spin-independent quantum dot Green's function  in spin space, $\textbf{g}=g\sigma^0$, where $\sigma^0$ is the identity matrix. It is given by the equation
\begin{align}
(i\partial_{t}-\varepsilon)\textbf{g} (t,t')=&\delta(t-t')+\int \boldsymbol{\Sigma} (t,\tau)\textbf{g} (\tau,t')d\tau,
\end{align}
where the self-energy is $\boldsymbol{\Sigma} = \Sigma\sigma^0$.
For the full derivation and solution to the Green's function, see Ref. \cite{Hammar2016}.

\subsection{Spin equation of motion}
Next, we consider the spin dynamics of the magnetic moment of the molecule and connect it to the phenomenological LLG equation.
The LLG equation in its extended form is defined as \cite{PhysRevLett.108.057204}
\begin{align}
\mathbf{\dot{S}}=\mathbf{S}\times(-\gamma\mathbf{B}^{\mathrm{eff}}+\mathbb{\hat{G}}\mathbf{\dot{S}}+\mathbb{\hat{I}}\mathbf{\ddot{S}}),
\label{extendedLLG}
\end{align}
where $\mathbf{B}^{\mathrm{eff}}$, $\mathbb{\hat{G}}$ and $\mathbb{\hat{I}}$ is the effective magnetic field, Gilbert damping and the moment of inertia tensor, respectively.
The moment of inertia term, $\mathbb{\hat{I}}$, has here been added in comparison to the conventional LLG equation, as there has been suggestions of its importance to short-time dynamics \cite{PhysRevLett.108.057204,PhysRevB.92.104403}.

The shortcomings of the conventional LLG equation is that its parameters are both constant and local.
For non-equilibrium conditions, this approach does not give the full picture as it fails to capture variations through space and time.
Therefore a microscopic approach is needed.
Using non-equilibrium conditions, we can define an effective action of the spin in an electronic environment that mediates the interactions between the spins in both time and space \cite{PhysRevLett.92.107001,PhysRevLett.108.057204,PhysRevLett.113.257201}.
If we integrate out the fermionic degrees of freedom and minimize the action, assuming a single classical spin, we can derive a generalized SEOM, given by
\begin{align}
\dot{\mathbf{S}} (t) = &\mathbf{S} (t)\times\left( -g\mu_{B}\mathbf{B}^{\mathrm{eff}}_{0} (t)+\frac{1}{e}\int \mathbb{J} (t,t')\cdot\mathbf{S} (t')dt'\right).
\label{spinequationofmotion}
\end{align}
Here, $\mathbf{B}^{\mathrm{eff}}_{0} (t)$ is the effective magnetic field acting on the spin
and $\mathbb{J} (t,t')$ is the dynamical exchange coupling tensor between spins at different times. The effective magnetic field is defined as
\begin{align}
\mathbf{B}^{\mathrm{eff}}_{0} (t)=&
\textbf{B}^{\mathrm{ext}}
-
\frac{1}{eg\mu_{B}}\int\textbf{j} (t,t')dt',
\end{align}
where the first term is the external magnetic field and the second term is the internal magnetic field due to the electron flow.
In the derivation, we assumed a classical spin of constant length, and ignored quantum fluctuations \cite{Hammar2016}.
While Eq. \eqref{extendedLLG} is an ordinary differential equation, Eq. \eqref{spinequationofmotion} is an integro-differential equation.
Hence, while the former is a simple instant approximation, the latter provides a description based on the whole past evolution of the spin.
As we shall see below, this difference has far reaching consequences in the final result.

We retain Eq. \eqref{extendedLLG} from Eq. \eqref{spinequationofmotion} by assuming that \textbf{S} is slowly varying with time, $\mathbf{S} (t') \approx \mathbf{S} (t) - (t-t')\mathbf{\dot{S}} (t) + (t-t')^2\mathbf{\ddot{S}} (t)/2$, which leads to
\begin{align}
&\frac{1}{e}\int \mathbb{J} (t,t')\cdot\mathbf{S} (t')dt' \approx \frac{1}{e} \left(\int \mathbb{J} (t,t')dt'\mathbf{S} (t)\right.\nonumber
\\
&\left.- \int\mathbb{J} (t,t')(t-t')dt'\mathbf{\dot{S}} (t) + \int \mathbb{J} (t,t')(t-t')^2 dt'\mathbf{\ddot{S}} (t)/2\right).
\label{taylorexpansion}
\end{align}
Here, the first term adds a contribution to the effective magnetic field, the second term corresponds to the Gilbert damping and the third term to the moment of inertia.
In the form of the LLG equation we identify the renormalized effective magnetic field
\begin{equation}
\mathbf{B}^{\mathrm{eff}} (t) = \textbf{B}^{\mathrm{ext}}
-
\frac{1}{eg\mu_{B}}\left(\int\textbf{j} (t,t')dt' + \int \mathbb{J} (t,t')dt'\mathbf{S} (t)\right)\label{timedependentfield}
\end{equation}
the damping tensor
\begin{equation}
\mathbb{\hat{G}} (t) = -\frac{1}{e}\int \mathbb{J} (t,t')(t-t')dt',\label{timedependentdamping}
\end{equation}
and the moment of inertia tensor
\begin{equation}
\mathbb{\hat{I}} (t) = \frac{1}{2e}\int \mathbb{J} (t,t')(t-t')^{2}dt'.\label{timedependentinertia}
\end{equation}
This is still more general than the conventional LLG equation, as the parameters depend on the time evolution of the charge and spin background through a memory kernel.

\subsection{Exchange coupling}

The internal magnetic field due to the electron flow is defined as
\begin{align}
\textbf{j} (t,t')=&
iev\theta(t-t')\av{\com{\text{s}^{(0)} (t)}{\mathbf{s} (t')}}.
\end{align}
Here, $\text{s}^{(0)}=\sum_{\sigma}\varepsilon_{\sigma}d_{\sigma}^{\dagger}d_{\sigma}/2=\psi^{\dagger}\boldsymbol{\epsilon}\psi/2$ is the on-site energy distribution, where $\boldsymbol{\epsilon}={\rm diag}\lbrace \varepsilon_\uparrow\ \varepsilon_\downarrow\rbrace$.
This two-electron Green's function is approximated by a decoupling into single electron Green's functions according to
\begin{eqnarray}
\mathbf{j} (t,t')&\approx&
iev\theta(t-t'){\rm sp}\boldsymbol{\epsilon}
\Bigl(
\mathbf{G}^{<} (t',t)\mathbf{\boldsymbol{\sigma}G}^{>} (t,t')
\nonumber\\&&
-\mathbf{G}^{>} (t',t)\mathbf{\boldsymbol{\sigma}G}^{<} (t,t')
\Bigr),
\label{currentMF}
\end{eqnarray}
where ${\rm sp}$ denotes the trace over spin 1/2 space.

The current $\mathbb{J} (t,t')=i2ev^2\theta(t-t')\av{\com{\bfs(t)}{\bfs(t')}}$ is the electron spin-spin correlation function which mediates the interactions between the localized magnetic moment at times $t$ and $t'$.
Analogously as the internal magnetic field, we decouple this two-electron Green's function according to
\begin{align}
\mathbb{J} (t,t')\approx&
\frac{ie}{2}v^{2}\theta(t-t'){\rm sp}\boldsymbol{\sigma}
\Bigl(
\mathbf{G}^{<} (t',t)\mathbf{\boldsymbol{\sigma}G}^{>} (t,t')
\nonumber\\&
-\mathbf{G}^{>} (t',t)\mathbf{\boldsymbol{\sigma}G}^{<} (t,t')
\Bigr).
\end{align}
This current mediated interaction can be decomposed into an isotropic Heisenberg interaction, $\text{J}_{H}$, and Ising, $\mathbb{J}_I$, and anisotropic Dzyaloshinski-Moriya (DM), ${\bf J}_{D}$, interactions.
This can be seen from the product ${\bf S}\cdot\mathbb{J}\cdot{\bf S}$, which is the corresponding contribution in the effective spin model \cite{PhysRevLett.113.257201} to ${\bf S} (t)\times\mathbb{J} (t,t')\cdot{\bf S} (t')$ in the generalized SEOM \cite{Hammar2016}.
This leads to that we can partition the exchange interaction in the generalized SEOM into
\begin{align}
\mathbf{S} (t)\times\mathbb{J} (t,t')\cdot\mathbf{S} (t')=&
\text{J}_{H} (t,t')\mathbf{S} (t)\times\mathbf{S} (t')
\nonumber\\&
+\mathbf{S} (t)\times\mathbb{J}_{I} (t,t')\cdot\mathbf{S} (t')
\nonumber\\&
-\mathbf{S} (t)\times\left[\mathbf{J}_{D} (t,t')\times\mathbf{S} (t')\right]\label{partitionSEOM}
\end{align}
where $\text{J}_{H}$ is a scalar, $\mathbb{J}_I$ is a tensor and ${\bf J}_{D}$ is a vector.
Effectively this corresponds to the Hamiltonian
\begin{align}
\mathscr{H}=& \mathbf{S}\cdot \left(\text{J}_{H}\mathbf{S}
+\mathbb{J}_{I}\cdot\mathbf{S}+\mathbf{J}_{D}\times\mathbf{S}\right).
\end{align}
In the adiabatic approximation, this gives the damping
\begin{align}
\mathbf{S}\times\mathbf{\hat{G}}\cdot\mathbf{\dot{S}}= &\hat{G} (\text{J}_{H})\mathbf{S}\times\mathbf{\dot{S}}
+\mathbf{S}\times\hat{G} (\mathbb{J}_{I})\cdot\mathbf{\dot{S}}
\nonumber\\&
-\mathbf{S}\times(\hat{G} (\mathbf{J}_{D})\times\mathbf{\dot{S}}),
\end{align}
and moment of inertia
\begin{align}
\mathbf{S}\times\mathbf{\hat{I}}\cdot\mathbf{\ddot{S}}=&\hat{I} (\text{J}_{H})\mathbf{S}\times\mathbf{\ddot{S}}
+\mathbf{S}\times\hat{I} (\mathbb{J}_{I})\cdot\mathbf{\ddot{S}}
-\mathbf{S}\times(\hat{I} (\mathbf{J}_{D})\times\mathbf{\ddot{S}}).
\end{align}
It is important to note here, that the exchange coupling mediates both isotropic and anisotropic terms in both the effective magnetic field, the Gilbert damping and the moment of inertia tensor in the framework of the LLG equation.
In its general form, this is also mediated in time, since the electronic structure depends on the spin dynamics.

The above treatment incorporates a current driven spin transfer torque, i.e., $\mathbf{S}\times\left[\mathbf{I}_{S}\times\mathbf{S}\right]$, where $\mathbf{I}_{S}$ is the spin current through the system \cite{Ralph20081190}.
This is included in the DM interaction, last term in Eq. \eqref{partitionSEOM}.
The DM interaction can be interpreted as a current through the system and describes a general form of spin current mediated interaction.
It is analogues to the spin transfer torque term found in similar treatments of ouf-of-equilibrium spin systems \cite{Ludwig2017, Chudnovskiy2008}.

\subsection{Stationary limit}
In the stationary limit, the exchange coupling and the parameters of the equation of motion simplifies further. Ignoring the moment of inertia term, our equation of motion becomes
\begin{align}
\dot{\mathbf{S}} (t) = &\mathbf{S} (t)\times\left( -g\mu_{B}\mathbf{B}^{\mathrm{eff}}+ \mathbb{\hat{G}}\cdot\mathbf{\dot{S}} (t)\right). \label{stationaryLLG}
\end{align}
The electron spin-spin correlation function can be Fourier transformed into energy space
\begin{align}
\mathbb{J} (\epsilon) = &
\frac{e}{2}v^{2}\int\frac{1}{\omega+\epsilon-\omega'+i\delta}{\rm sp}\boldsymbol{\sigma}
\Bigl(
\mathbf{G}^{<} (\omega)\mathbf{\boldsymbol{\sigma}G}^{>} (\omega')
\nonumber\\&
-\mathbf{G}^{>} (\omega)\mathbf{\boldsymbol{\sigma}G}^{<} (\omega')
\Bigr)\frac{d\omega}{2\pi}\frac{d\omega'}{2\pi},
\end{align}
where we used the fact that the Green's function can be rewritten as $G(t,t') = G(t-t')$ in the stationary limit.
This can analogously to the time-dependent case be decomposed into a Heisenberg, Ising and DM term, as done in Ref. \cite{PhysRevLett.113.257201,Hammar2016}.

The Gilbert damping can in the stationary limit be derived from \cite{PhysRevLett.108.057204}
\begin{align}
\mathbb{\hat{G}} = & -\frac{1}{e}\int \mathbb{J} (t,t')(t-t')dt' = - \frac{1}{e}lim_{\epsilon \rightarrow 0}i\partial_{\epsilon} \mathbb{J} (\epsilon), \nonumber\\
= & -\frac{1}{2}v^{2}\mathrm{Im}\int\frac{1}{(\omega-\omega'+i\delta)^{2}}{\rm sp}\boldsymbol{\sigma}
\Bigl(
\mathbf{G}^{<} (\omega)\mathbf{\boldsymbol{\sigma}G}^{>} (\omega')
\nonumber\\&
-\mathbf{G}^{>} (\omega)\mathbf{\boldsymbol{\sigma}G}^{<} (\omega')
\Bigr)\frac{d\omega}{2\pi}\frac{d\omega'}{2\pi}.\label{constantdamping}
\end{align}

\section{Results}
\begin{figure*}[t]
	\centering
	\includegraphics[width=\textwidth]{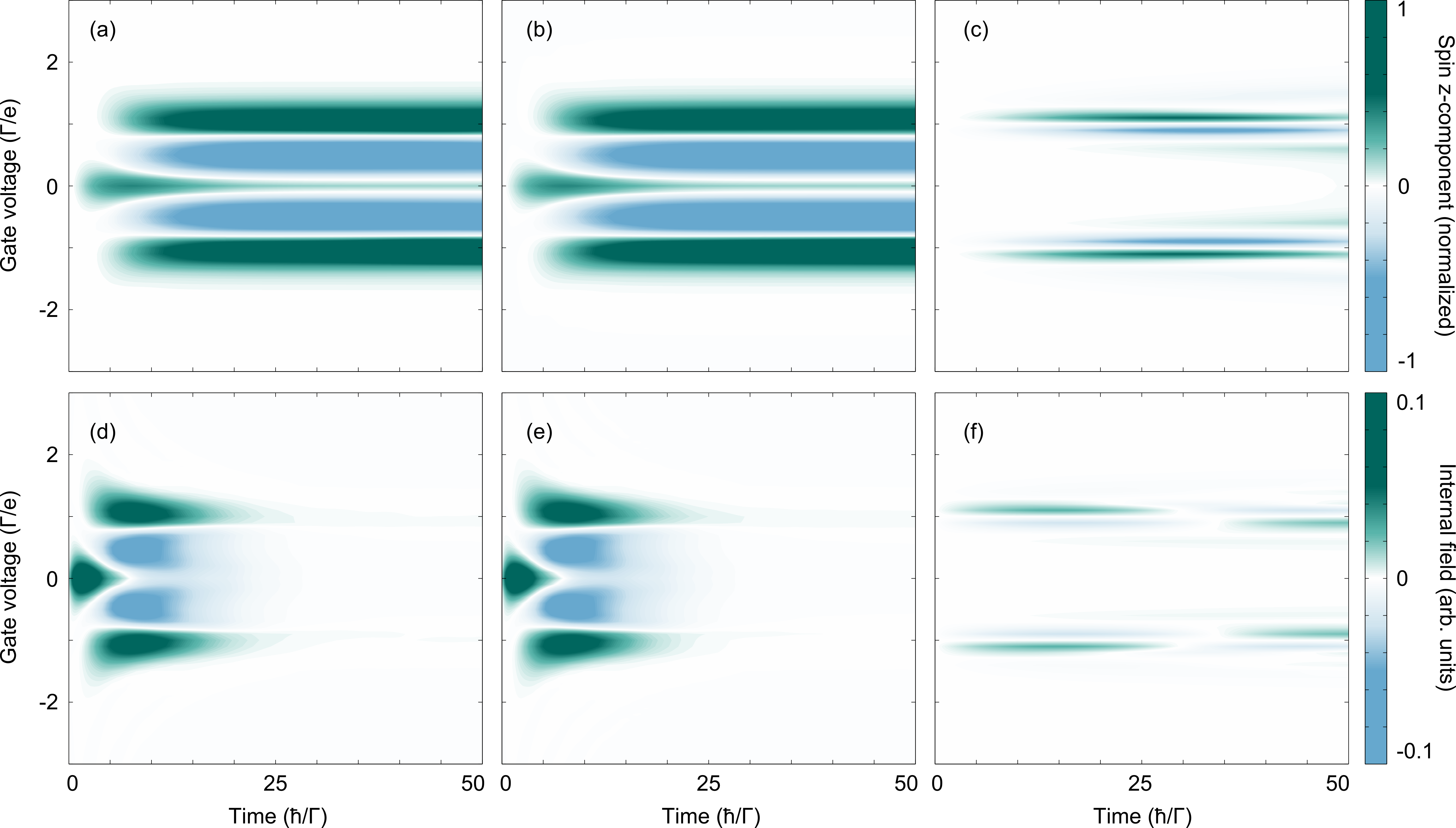}
	\caption{Evolution of $S_{z}$ for different gate voltage for (a) the generalized SEOM, (b) the tdLLG solution and (c) the stationary LLG solution.  In the bottom row the effect from the internal field from the charge flow in the quantum dot, $S\times\int\textbf{j} (t,t')dt'$, is shown for (d) the generalized SEOM,  (e) the tdLLG solution and (f) the stationary LLG solution. Here,  eV = $20\Gamma/3$, $v = \Gamma/3$, B = $1.158\cdot10^{-4}$ $\Gamma/\text{g}\mu_{B}$ and T = $8.617\cdot10^{-2}$ $\Gamma/\text{k}_{B}$.}
	\label{DiffSzsolution}
\end{figure*}

In the following Eq. \eqref{spinequationofmotion} is referred to as the generalized SEOM, while the time-dependent LLG equation of motion, Eq. \eqref{extendedLLG}, with parameters given by Eqs. \eqref{timedependentfield} -- \eqref{timedependentinertia}, is referred to as the tdLLG, while the constant LLG equation of motion, with parameters given by Eqs. \eqref{stationaryLLG} and \eqref{constantdamping}, is denoted LLG.

We test the different approximation schemes by comparing them with the results of the generalized SEOM for different exchange couplings.
In order to study the transient regime, the system has an abrupt on-set of a voltage bias and the exchange interaction at time $t_{0}$.

We begin by considering the low coupling regime for symmetrically coupled leads, i.e., $ v < \Gamma_{L} = \Gamma_{R} = \Gamma$.
We identify this regime being at the star in Fig. \ref{solutiondiagram}, with dynamics at about 1 $\hbar/\Gamma$ and low exchange in relation with $\Gamma$.
In Fig. \ref{DiffSzsolution} (a) the solution of the full equation of motion is shown together with the solution of the tdLLG equation, Fig. \ref{DiffSzsolution} (b), and the solution for the LLG with constant parameters, Fig. \ref{DiffSzsolution} (c).
The plots show the $S_{z}$ component of the local spin and its evolution due to the sudden on-set.
For the generalized SEOM, Fig. \ref{DiffSzsolution} (a), there are sudden changes in the transient regime which evolves into a stable stationary solution.
These main features are reproduced within the tdLLG equation, see Fig. \ref{DiffSzsolution} (b).

However, in the case of constant parameters, Fig. \ref{DiffSzsolution} (c), the solution clearly differs from the other two.
The origin of this difference can be observed in the internal field, $\int\textbf{j} (t,t')dt'$, shown in Fig. \ref{DiffSzsolution} (d)--(f) for the three different schemes.
The internal field changes rapidly after the on-set, Fig. \ref{DiffSzsolution} (d)--(e), while it increases adiabatically for the static parameters, Fig. \ref{DiffSzsolution} (f) (do also note the difference of the size of the field by an order of magnitude).
The differences in the stationary limit solutions vividly illustrate that the time-dependence of internal magnetic and exchange fields has a vital influence on the dynamics, also in the far future.
Thus, the failure of the conventional LLG-equation shows that initial transients changes the long-time characteristics of the system.
It also shows the importance of treating the exchange dynamically in order to incorporate fast changes in the system.
Similar results have been achieved in studies on dynamical exchange splitting \cite{Mueller2013}.

\begin{figure*}[t]
	\centering
	\includegraphics[width=\textwidth]{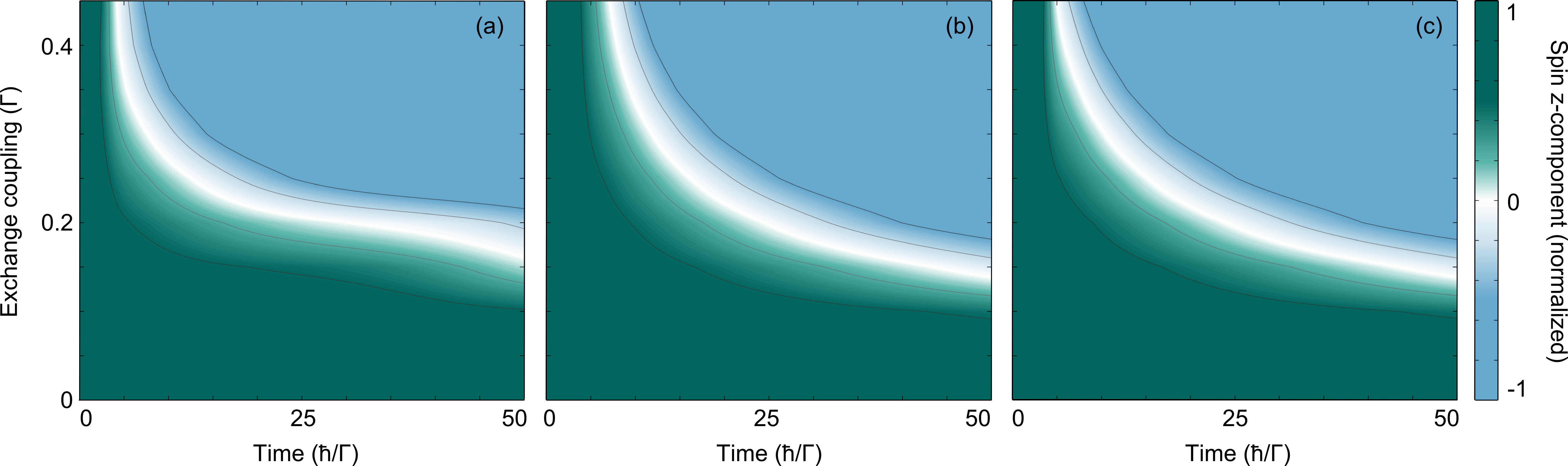}
	\caption{Evolution of $S_{z}$ for different exchange coupling $v$ for (a) the generalized SEOM, (b) the tdLLG solution with damping and (c) the tdLLG solution with both damping and moment of intertia. The shaded region indicates the transition from spin up to spin down. The lines indicates $S_{z} = \pm 0.25$ and $\pm 0.5$ respectively. Here, eV = $2\Gamma$, B = $0.1158$ $\Gamma/\text{g}\mu_{B}$ and T = $8.617\cdot10^{-2}$ $\Gamma/\text{k}_{B}$.}
	\label{DiffExchange}
\end{figure*}

Thus far, we have only included damping for the tdLLG solution.
By increasing the exchange coupling, this approach fails to reproduce the dynamics properly.
Comparing the generalized SEOM, Fig. \ref{DiffExchange} (a), with the tdLLG solution with only damping, Fig. \ref{DiffExchange} (b), and with both damping and moment of inertia, Fig. \ref{DiffExchange} (c), it is clear that the tdLLG approach fails to capture the full dynamics as the exchange coupling increase.
Comparing the generalized SEOM, Fig. \ref{DiffExchange} (a), with the tdLLG with only damping, \ref{DiffExchange} (b), it can be noticed that the latter approach neither captures the fast dynamics for $v\sim0.4\Gamma$ nor the scaling behavior of the dynamics at about $v\sim0.15\Gamma$--$0.3\Gamma$.
In the tdLLG the impact of the local exchange $v$ goes like $v^4$, while in the generalized SEOM the scaling is non-linear in $v^4$.
By including the moment of inertia in the tdLLG, Fig. \ref{DiffExchange} (c), the fast dynamics for higher exchange is partially recreated, while the scaling in the regime $v\sim0.15\Gamma$--$0.3\Gamma$ is not. We can, therefore, conclude that while some of the dynamics can be captured by higher order terms in the tdLLG, inclusion of the the history is quite necessary in order to retain the full dynamics of the spin. This shows the significance of considering non-Markovian and non-linear effects in the system.

Upon increasing the exchange coupling $v$ beyond what is considered in Fig. \ref{DiffExchange}, the applicability of all computational schemes discussed in this paper becomes questionable, as the limits of the different approaches are being reached.
We obtain numerical stability of the generalized SEOM the exchange coupling up to about $v\sim 0.7\Gamma$ for $\Gamma = 1$ meV.
For higher values of the exchange coupling the dynamics is faster than $0.1\hbar/\Gamma$ at which time scales the validity the Born-Oppenheimer approximation becomes dubious.
Using the tdLLG approximation under the same conditions, only $v\sim0.5\Gamma$ can be reached. Above that threshold the numerical solution breaks down as the indirect interaction diverges. Thus, the generalized SEOM enables simulations the dynamics of more strongly coupled systems than the tdLLG approach. We conjecture that for stronger local exchange coupling $v$, the appropriate spin dynamics has to be approached from an Anderson model perspective since the localized character of the spin cannot be justified. 

\section{Conclusion}
In conclusion we have compared three different approximation schemes for treating the transient spin dynamics of a magnetic molecule.
The results are summarized in Fig. \ref{solutiondiagram}.
They show that conventional LLG with constant parameters does not capture the fast dynamics in the system, while the tdLLG fails to capture the strongly coupled regime and fast dynamics.
In fact, using constant parameters may even lead to a completely different solution in the stationary limit, an effect which is well-known to be a risk in non-linear dynamics.
Therefore, inclusion of the full history in the time-evolution is necessary when approaching fast dynamics.
Using a generalized SEOM, we can incorporate both the changes in the electronic background and in the localized spin moment, and thereby treat faster dynamics.
While our study has been restricted to a single molecule, we believe that our results have implications in larger nanostructures and, hence, the interpretations and validity of spin dynamics using ab intio methods.

\acknowledgments
We want to thank J. D. Vasquez Jaramillo and K. Bj\"ornson for fruitful discussions about the work. Financial support from Vetenskapsr\aa det is acknowledged. This work is part of the CINT User Proposal \# U2015A0056.

\end{document}